\documentclass{aa}
\usepackage{txfonts}
\usepackage{graphicx}
\usepackage{natbib}
\usepackage[figuresright]{rotating}
\bibpunct{(}{)}{;}{a}{}{,}
\begin{document}

\title{Abundances of Mn, Co and Eu in a sample of 20 F--G disk stars: the influence
of hyperfine structure splitting
\thanks{Based on observations collected
at the European Southern Observatory, La Silla, Chile, under the
ESO programs and the ESO-Observat\'orio Nacional, Brazil,
agreement.}\fnmsep\thanks{Full Tables~\ref{tab:Mn_abundances} and
3, which contain line-by-line Mn and Co abun\-dances
(re\-spec\-tive\-ly), are only available in electronic form at the
CDS via anonymous ftp to {\tt cdsarc.u-strasbg.fr (130.79.128.5)}
or via {\tt http://cdsweb.u-strasbg.fr/cgi-bin/qcat?}.}}

\titlerunning{Abundances of Mn, Co and Eu: the influence
of hyperfine structure splitting}

\author{E.F. del Peloso\inst{1}\fnmsep\inst{2} \and K. Cunha\inst{1} \and L. da Silva\inst{1} \and
G.F. Porto de Mello\inst{2}}

\offprints{E.F. del Peloso}

\institute{Observat\'orio Nacional/MCT, Rua General Jos\'e
Cristino
77, 20921-400 Rio de Janeiro, RJ, Brazil\\
\email{epeloso@on.br, katia@on.br, licio@on.br} \and
Observat\'orio do Valongo/UFRJ, Ladeira
do Pedro Ant\^onio 43, 20080-090 Rio de Janeiro, RJ, Brazil\\
\email{gustavo@ov.ufrj.br}}

\date{Received 16 Mars 2005 / Accepted 22 June 2005}

\abstract{We present Mn, Co and Eu abundances for a sample of 20
disk F and G dwarfs and subgiants with metallicities in the range
$-0.8\le\mathrm{[Fe/H]}\le+0.3$. We investigate the influence of
hyperfine structure (HFS) on the derived abundances of Mn and Co
by using HFS data from different sources in the literature, as
well as calculated HFS from interaction factors $A$ and $B$. Eu
abundances were obtained from spectral synthesis of one
\ion{Eu}{ii} line that takes into account HFS from a series of
recent laboratory measurements. For the lines analyzed in this
study, we find that for manganese, the differences between
abundances obtained with different HFSs are no larger than
0.10~dex. Our cobalt abundances are even less sensitive to the
choice of HFS than Mn, presenting a 0.07~dex maximum difference
between determinations with different HFSs. However, the cobalt
HFS data from different sources are significantly different. Our
abundance results for Mn offer an independent confirmation of the
results from \citet{prochaska&mcwilliam00}, who favour type~Ia
supernovae as the main nucleosynthesis site of Mn production, in
contrast to trends of Mn versus metallicity previously reported in
the literature. For Co, we obtain $\mathrm{[Co/Fe]}\sim0.0$ in the
range $-0.3<\mathrm{[Fe/H]}<+0.3$ and [Co/Fe] rising to a level of
$+0.2$ when [Fe/H] decreases from $-0.3$ to $-0.8$, in
disagreement with recent results in the literature. The observed
discrepancies may be attributed to the lack of HFS in the works we
used for comparison. Our results for Eu are in accordance with
low-mass type~II supernovae being the main site of the
\emph{r}-process nucleosynthesis. }

\maketitle

\section{Introduction}

Accurate stellar abundance determinations are fundamental for
numerous astrophysical studies, like those of Galactic and stellar
structure, evolution, and nucleosynthesis. In order to study the
behavior of certain elements like Mn, Co, and Eu with metallicity,
it is a well-known fact that it is crucial to consider hyperfine
structure (HFS) splitting in the calculations using strong lines,
because otherwise the computed abundances are bound to be
erroneous. Moreover, abundance results computed adopting HFSs from
different sources can produce trends with metallicity that are
significantly different, as shown, for example, by \citet[\space
PM00]{prochaska&mcwilliam00}.

PM00 investigated the importance of HFS on abundance
determinations of Mn and Sc. In their study, they find that an
incorrect treatment for HFS can lead to abundances that are
significantly in error. For Mn, in particular, they discuss the
results of \citet[\space NCSZ00]{nissenetal00}, who used the HFS
components published by \citet[\space S85]{steffen85} which, in
turn, are based on the work of \citet{biehl76}. PM00 pointed out
that S85 grouped together nearby hfs components and applied old,
inaccurate splitting constants, and that such simplifications
introduce significant errors in the obtained Mn abundances,
producing spurious abundance trends with metallicity.

The purpose of this work is to investigate the behavior of Mn, Co,
and Eu from a sample of 20 F--G dwarfs and subgiants with
metallicities typical of the Galactic disk (in the range
$-0.8\le\mathrm{[Fe/H]}\le+0.3$), focussing on the evaluation of
the influence of HFS in the abundance determinations and abundance
trends. Concerning the latter, the main questions that we seek to
answer 1)~How large are the differences between abundances
obtained using HFS components from different sources? and 2)~How
large are the inaccuracies introduced when simplifications like
the grouping together of close-by components are used? In order to
accomplish this goal, two sets of Mn and Co abundances were
calculated for our sample: one with the HFS data from S85, and
another with HFS data from R.L. Kurucz's website (hereafter
referred to as KLL) For the \ion{Co}{i} lines, two additional sets
of calculations were also obtained: one without HFS and one using
HFSs calculated by us with interaction factors $A$ and $B$ taken
from the literature ($A$ and $B$ are the magnetic dipole and
electric quadrupole coupling constants, respectively). In
addition, Eu abundances for the sample stars (from
\citealt{delpelosoetal05a}) which were derived adopting HFS's
calculated with interaction factors from the literature, in a
similar manner as for Co I lines, will be discussed and compared
to abundance results from the literature.

\section{Sample selection, observations, data reduction and atmospheric
parameter determination}

The detailed description of sample selection, observations, data
reduction, and atmospheric parameter determination is given in
\citet{delpelosoetal05a}; in what follows, we provide here only a
brief overview of these topics.

The sample was originally selected to determine the age of the
Galactic thin disk through Th/Eu nucleocosmochronology. It is
composed of 20 dwarfs and subgiants of F5 to G8 MK spectral types
with $-0.8\le\mathrm{[Fe/H]}\le+0.3$, located less than 40~pc
away.

All objects were observed with the Fiber-fed Extended Range
Optical Spectrograph \citep[FEROS;\ ][]{kauferetal99} coupled to
the 1.52~m European Southern Observatory (ESO) telescope, as a
part of the ESO-Observat\'orio Nacional, Brazil, agreement. The
obtained spectra have high nominal resolving power (R~=~48\,000),
signal-to-noise ratio ($\mbox{S/N}\ge300$ in the visible) and
coverage (3500~\AA\ to 9200~\AA\ spread over 39~echelle orders).
Additional observations, centered around the \ion{Eu}{ii} line at
4129.72~\AA, were carried out with the Coud\' e \'Echelle
Spectrograph (CES) fiber-fed by ESO's Coud\'e Auxiliary Telescope
(CAT). The obtained spectra have high nominal resolving power
(R~=~50\,000) and signal-to-noise ratio ($\mbox{S/N}\sim300$);
coverage is 18~\AA.

A set of homogeneous, self-consistent atmospheric parameters was
determined for the sample stars. Effective temperatures were
determined from photometric calibrations and H$\alpha$ profile
fitting; surface gravities were obtained from $T_{\mathrm{eff}}$,
stellar masses and luminosities; microturbulence velocities and
metallicities were obtained from detailed, differential
spectroscopic analysis, relative to the Sun, using equivalent
widths (EWs) of \ion{Fe}{i} and \ion{Fe}{ii} lines.

\section{Abundance determinations}

\subsection{Manganese and cobalt}
\label{sec:mn_co_abund}

Mn and Co abundances were determined using EWs of 6~\ion{Mn}{i}
and 8~\ion{Co}{i} lines measured in FEROS spectra. As mentioned
above, two sets of abundance calculations for Mn and Co were
obtained, with HFS data from S85 and KLL. For the \ion{Co}{i}
lines, two additional sets were also obtained, without HFS and
with HFS calculated with Casimir's equation \citep{casimir_63}:
\begin{displaymath}
W_F=W_J+\frac{AK}{2}+\frac{3BK(K+1)-4I(I+1)J(J+1)}{8I(2I-1)J(2J-1)},
\end{displaymath}
where $W_F$ is the energy of the hyperfine level, $W_J$ is the
energy of the fine-structure level of quantum number~$J$, $I$ is
the nuclear spin, $K$ is defined as
\begin{displaymath}
K=F(F+1)-I(I+1)-J(J+1),
\end{displaymath}
and $F$ is the quantum number associated with the total angular
momentum of the atom,
\begin{displaymath}
F=I+J;I+J-1;\ldots;|I-J|.
\end{displaymath}
HFS transitions are governed by the following selection rules:
$\Delta F=0;\pm1$, but not $F=0\leftrightarrow F'=0$.

The energies of the fine-structure levels were taken from
\citet{pickering&thorne96}, and the $A$ and $B$ constants from
\citet{guthohrlein&keller90} and \citet{pickering96}. Intensities
of the components were obtained by distributing the total $\log
gf$ values according to the relative weights tabulated in 1933 by
White \& Eliason \citep{condon&shortley67}.  The Co HFSs derived
are presented in Table~\ref{tab:Co_hfs_1}. Solar $\log gf$ values
were used for all Mn and Co lines. These were determined by
forcing the abundances obtained with solar spectra to match those
from \citealt{grevesse&sauval98}
($\log\varepsilon(\mbox{Mn})=5.39$ and
$\log\varepsilon(\mbox{Co})=4.92$).

The adopted metallicities ([Fe/H]) were taken from
\citet{delpelosoetal05a}. Table~\ref{tab:Mn_abundances} presents a
sample of the [Mn/H] results on a line-by-line basis. Its complete
content, composed of the abundances of all measured lines, for all
sample stars, obtained with all HFS sources employed, is only
available in electronic form at the CDS. Column~1 lists the HD
number of the object. Subsequent columns present the [Mn/H]
abundance ratios. Table~3, which contains the line-by-line [Co/H]
abundance ratios, is formatted in this same manner and is also
only available electronically.

\begin{table*}
\caption[]{HFSs of all \ion{Co}{i} lines calculated in this study}
\label{tab:Co_hfs_1}\centering
\begin{tabular}{ c r @{.} l c r @{.} l c r @{.} l c r @{.} l} \hline \hline
\multicolumn{3}{c}{4749.612~\AA} &
\multicolumn{3}{c}{5212.691~\AA} &
\multicolumn{3}{c}{5280.629~\AA} &
\multicolumn{3}{c}{5301.047~\AA}\\
$\lambda$~(\AA) & \multicolumn{2}{c}{W~(\%)} & $\lambda$~(\AA) &
\multicolumn{2}{c}{W~(\%)} & $\lambda$~(\AA) &
\multicolumn{2}{c}{W~(\%)} &  $\lambda$~(\AA) &
\multicolumn{2}{c}{W~(\%)}\\
\hline 4749.616 &19 & 850 & 5212.595 &  1 & 732 & 5280.562 &  0 &
256 & 5301.014 & 4 & 517\\%1
4749.634 &16 & 321 & 5212.622 & 2  & 842 & 5280.569 &  2 & 088 &
5301.023 &22 & 584\\%2
4749.650 &13 & 266 & 5212.646 &  3 & 409 & 5280.586 &  0 & 426 &
5301.032 & 6 & 836\\%3
4749.665 & 10 & 635 & 5212.653 & 20 & 168 & 5280.591 &  3 & 386 &
 5301.040 & 11 & 529\\%4
4749.678 &  8 & 370 & 5212.668 & 3 & 513 & 5280.606 & 0 & 533 &
5301.046 &  7 & 393\\%5
4749.684 & 1  & 408 &5212.673&14&745          & 5280.608 &21 & 318
&  5301.049 &  4 & 517\\%6
4749.691 &  6 & 498 &5212.687&3&145           & 5280.609 & 4 & 025
&  5301.054 &  4 & 517\\%7
4749.694 & 2  & 343 &  5212.691 & 10 & 487    & 5280.623 & 0 & 533
&  5301.058 &  6 & 365\\%8
4749.702 & 4 & 940 & 5212.704 & 2 & 498 &  5280.625 & 20& 773
 & 5301.062 & 6 & 836\\%9
4749.703 & 2 & 817 & 5212.705 & 7 & 255 & 5280.637 & 4 & 007
 & 5301.064 & 0 & 859\\%10
 4749.710 & 2 & 916 & 5212.717 & 4 & 917 & 5280.638 &12 & 786
 &5301.068 & 4 & 016\\%11
4749.712 & 3 & 748 & 5212.726 & 3 & 347 & 5280.647 & 2 & 876
 & 5301.071 & 7 & 438\\%12
4749.716 & 2 & 702 & 5212.732 & 1 & 732 & 5280.649 & 9 & 500
 & 5301.076 & 2 & 212\\%13
4749.721 & 2 & 161 & 5212.733 & 2 & 442 & 5280.653 & 1 & 831
 & 5301.077 & 6 & 365\\%14
4749.724 & 1 & 330 & 5212.738 & 2 & 359 & 5280.656 & 6 & 804
 & 5301.080 & 4 & 016\\%15
4749.743 & 0 & 139 & 5212.742 & 2 & 842 & 5280.661 & 6 & 155
&  & \multicolumn{2}{c}{}\\%16
4749.746 & 0 & 218 & 5212.749 & 3 & 409 & 5280.662 & 2 & 702
&  & \multicolumn{2}{c}{}\\%17
4749.747 & 0 & 119 & 5212.754 & 3 & 513 &  & \multicolumn{2}{c}{}
&  & \multicolumn{2}{c}{}\\%18
4749.748 & 0 & 218 & 5212.755 & 2 & 498 &  & \multicolumn{2}{c}{}
&  & \multicolumn{2}{c}{}\\%19
  & \multicolumn{2}{c}{} & 5212.756 & 3 & 145 &  &
\multicolumn{2}{c}{} & & \multicolumn{2}{c}{}\\%20
\end{tabular}

\vspace{0.1cm}

\begin{tabular}{c r @{.} l c r @{.} l  c r @{.} l c r @{.} l} \hline \hline
\multicolumn{3}{c}{5342.708~\AA} &
\multicolumn{3}{c}{5454.572~\AA} &
\multicolumn{3}{c}{5647.234~\AA} &
\multicolumn{3}{c}{6188.996~\AA}\\
$\lambda$~(\AA) & \multicolumn{2}{c}{W~(\%)} & $\lambda$~(\AA) &
\multicolumn{2}{c}{W~(\%)} & $\lambda$~(\AA) &
\multicolumn{2}{c}{W~(\%)} & $\lambda$~(\AA) &
\multicolumn{2}{c}{W~(\%)}\\
\hline 5342.699 & 6 & 071 & 5454.551 &  2 & 003 & 5647.212 & 31 &
657 & 6188.925 & 4 & 517\\%1
5342.700 &10 & 870 & 5454.553 &  3 & 941 & 5647.225 & 19 &
296 & 6188.938 & 22& 584\\%2
5342.702 & 13& 099 & 5454.555 &  4 & 061 & 5647.237 & 10 &
034 & 6188.968 & 6 & 836\\%3
5342.705 & 15& 676 & 5454.557 &  3 & 662 & 5647.243 &  7 &
473 & 6188.979 &11 & 529\\%4
5342.707 & 1 & 136 & 5454.560 &  2 & 888 & 5647.248 & 3  &
643 & 6188.991 & 4 & 517\\%5
5342.708 &18 & 545 & 5454.568 &  2 & 727 & 5647.251 & 10 &
826 & 6189.002 & 7 & 393\\%6
5342.710 & 1 & 876 & 5454.569 &  8 & 388 & 5647.258 &  1 &
139 & 6189.012 & 4 & 517\\%7
5342.712 & 21& 839 & 5454.571 & 12 & 124 & 5647.264 &  7 &
319 & 6189.023 & 6 & 836\\%8
5342.713 & 2 & 335 & 5454.572 & 17 & 047 & 5647.269 &  1 &
013 & 6189.030 & 6 & 365\\%9
5342.718 & 2 & 531 & 5454.575 &  23& 316 & 5647.272 &  2 &
719 & 6189.038 & 0 & 859\\%10
5342.719 & 0 & 065 & 5454.577 &  2 & 888 & 5647.274 &  4 &
881 & 6189.048 & 7 & 393\\%11
5342.722 & 2 & 400 & 5454.580 &  3 & 662 & & \multicolumn{2}{c}{}
 & 6189.052 & 4 & 016\\%12
5342.724 & 0 & 109 & 5454.583 &  4 & 061 & & \multicolumn{2}{c}{}
 & 6189.057 & 0 & 045\\%13
5342.728 & 1 & 964 & 5454.587 &  3 & 941 & & \multicolumn{2}{c}{}
 & 6189.064 & 6 & 365\\%14
5342.731 & 0 & 131 & 5454.592 &  3 & 286 & & \multicolumn{2}{c}{}
 & 6189.069 & 2 & 212\\%15
5342.734 & 1 & 178 & 5454.597 &  2 & 003 & & \multicolumn{2}{c}{}
 & 6189.075 & 4 & 016\\%16
5342.738 & 0 & 109 & & \multicolumn{2}{c}{}& &
\multicolumn{2}{c}{}
& & \multicolumn{2}{c}{}\\%17
5342.745 & 0 & 065& & \multicolumn{2}{c}{}& & \multicolumn{2}{c}{}& & \multicolumn{2}{c}{}\\%18
\hline\end{tabular}

\vspace{-0.3cm}

\begin{flushleft} {\hspace{3.3cm}Note: ``W'' stands for weight.}
\end{flushleft}
\end{table*}

\begin{table}
\caption[]{A sample of the [Mn/H] abundance ratios, line-by-line.
Abundances are presented for all HFS sources, for all measured
lines. The complete content of this table is only available in
electronic form at the CDS, along with a similar table for [Co/H]
abundance ratios. For a description of the columns, see text
(Sect.~\ref{sec:mn_co_abund}).} \label{tab:Mn_abundances}
\begin{tabular}{l c @{\hspace{1em}} c @{\hspace{1em}} c c c c}
\hline \hline  HD & 4739.113~\AA &&
5394.670~\AA & $\cdots$ & \multicolumn{2}{c}{5432.538~\AA}\\
\cline{2-2}\cline{4-4}\cline{6-7}&martin&&martin&&steffen&martin\\
 \hline
2151 & $-$0.08 && $-$0.12 & $\cdots$ & $-$0.14 & $-$0.13\\
9562 & +0.22 && +0.10 & $\cdots$ & +0.12 & +0.12\\
\multicolumn{1}{c}{$\vdots$} & $\vdots$ && $\vdots$ & $\vdots$ &
$\vdots$ &
$\vdots$\\
199\,288 & $-$0.75 && $-$0.78 & $\cdots$ & $-$0.79 & $-$0.78\\
203\,608 & $-$0.75 && $-$0.72 & $\cdots$ & -- & --\\
\hline
\end{tabular}
\vspace{0.1cm}

Note: Column labels indicate the source of the HFS: martin --
\citet{martinetal88}, steffen -- \citet{steffen85}, kurucz --
\citet{kurucz90}. \vfill
\end{table}

\subsection{Europium}

Eu abundances were taken from \citet{delpelosoetal05a}. They were
obtained from spectral synthesis of the \ion{Eu}{ii} line at
4129.72~\AA. HFS was calculated by us in exactly the same way as
for Co, using data from \citet{beckeretal93},
\citet{molleretal93}, \citet{villemoes&wang94}, and
\citet{brostrometal95}. Isotope shift was taken into account,
using data from \citet{brostrometal95} and the solar abundance
isotopic ratio
$\varepsilon(\mbox{\element[][151]{Eu}})/\varepsilon(\mbox{\element[][153]{Eu}})=1.00\pm0.29$
\citep{lawleretal01}. The complete HFSs obtained for both Eu
isotopes are presented in Table~\ref{tab:Eu_hfs}.

\begin{table}
\setcounter{table}{3} \caption[]{HFSs of the \ion{Eu}{ii} line.}
 \label{tab:Eu_hfs}
\begin{tabular}{ c c r @{.} l} \hline \hline
$\lambda_{151}$(\AA) & $\lambda_{153}$(\AA) &
\multicolumn{2}{c}{W}\\
\hline
4129.615 & 4129.695 & 0 & 923\\%1
4129.618 & 4129.698 & 2 & 792\\%2
4129.632 & 4129.702 & 1 & 462\\%3
4129.636 & 4129.705 & 3 & 170\\%4
4129.640 & 4129.708 & 0 & 922\\%5
4129.657 & 4129.712 & 1 & 664\\%6
4129.662 & 4129.715 & 4 & 285\\%7
4129.667 & 4129.719 & 1 & 462\\%8
4129.690 & 4129.727 & 1 & 531\\%9
4129.696 & 4129.730 &  6 & 053\\%10
4129.702 & 4129.733 & 1 & 663\\%11
4129.731 & 4129.747 & 0 & 993\\%12
4129.738 & 4129.748 &  8 & 590\\%13
4129.744 & 4129.751 & 1 & 531\\%14
4129.788 & 4129.773 & 11 & 966\\%15
4129.795 & 4129.774 & 0 & 993\\%16
\hline
\end{tabular}

Note: $\lambda_{151}$ and $\lambda_{153}$ are the central
wavelengths of components for $^{151}$Eu and $^{153}$Eu,
respectively. `W'' stands for the weight of each component and
each isotope.
\end{table}

We kept the $\log gf$ value fixed at the laboratory value provided
by \citet{komarovskii91} and derived a solar abundance using solar
spectrum. Abundances for the sample stars were obtained relative
to this solar value.

\section{Abundance results obtained with HFS from different sources}

\subsection{Manganese}
\label{sec:mn_line_by_line}

Of the 6 \ion{Mn}{i} lines used in this study, 4 of them
(5399.479~\AA, 5413.684~\AA, 5420.350~\AA, and 5432.548~\AA) have
HFS data available from both S85 and KLL. Note that KLL assembles
data from multiple sources; for the \ion{Mn}{i} lines studied
here, they come from \citet{martinetal88} and \citet{kurucz90}.
For these 4 lines, the structures of S85, although simplified by
the grouping together of close-by components, are very similar to
those from KLL. We note, however, that the HFSs for the three
\ion{Mn}{i} lines used in the studies of NCSZ00 and PM00
(6013.513~\AA, 6016.673~\AA, and 6021.819~\AA) are, on the other
hand, quite different in S85 and KLL. The differences in the HFSs
offer an explanation for the fact that our abundances obtained
with HFS from S85 do not match those from NCSZ00, although taken
from the same reference. This is because, while the HFSs from S85
for the six \ion{Mn}{i} lines used by us appear to have been
accurately calculated, those for the three lines used in NCSZ00
appear to contain important deviations when compared to KLL.

\begin{figure}
\resizebox{\hsize}{!}{\includegraphics*{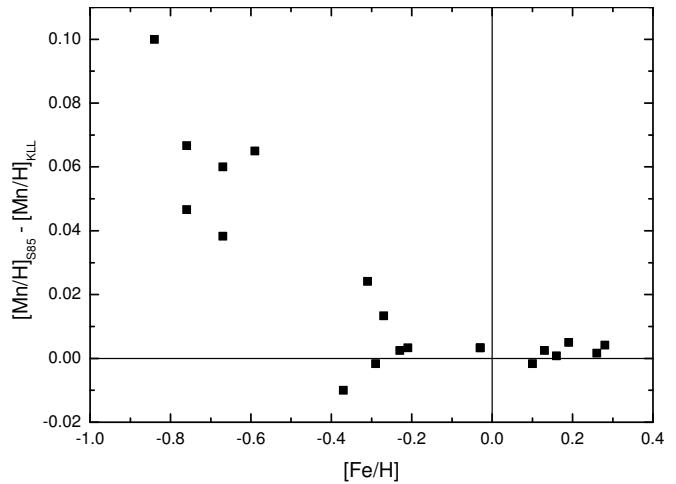}}
\caption{Diagram displaying the difference between average [Mn/H]
abundance ratios obtained with the HFSs from S85 and KLL, for our
sample stars.} \label{fig:mn_st_minus_mn_ku}
\end{figure}

In Figure~\ref{fig:mn_st_minus_mn_ku} we show the differences
between average [Mn/H] abundance ratios obtained for our sample
with the two adopted HFSs, from S85 and KLL. The abundances are
very similar at roughly solar metallicities and down to roughly
$\mathrm{[Fe/H]}=-0.2$, but the differences are larger at lower
metallicities, reaching a maximum value of 0.10~dex. This
dependance of the abundance differences with metallicity (although
reaching only the modest level of 0.10~dex) would create a
spurious trend in the run of Mn abundance with metallicity that is
only due to choice of HFS.

\subsection{Cobalt}

\begin{figure*}
\centering
\includegraphics*[width=16.54cm]{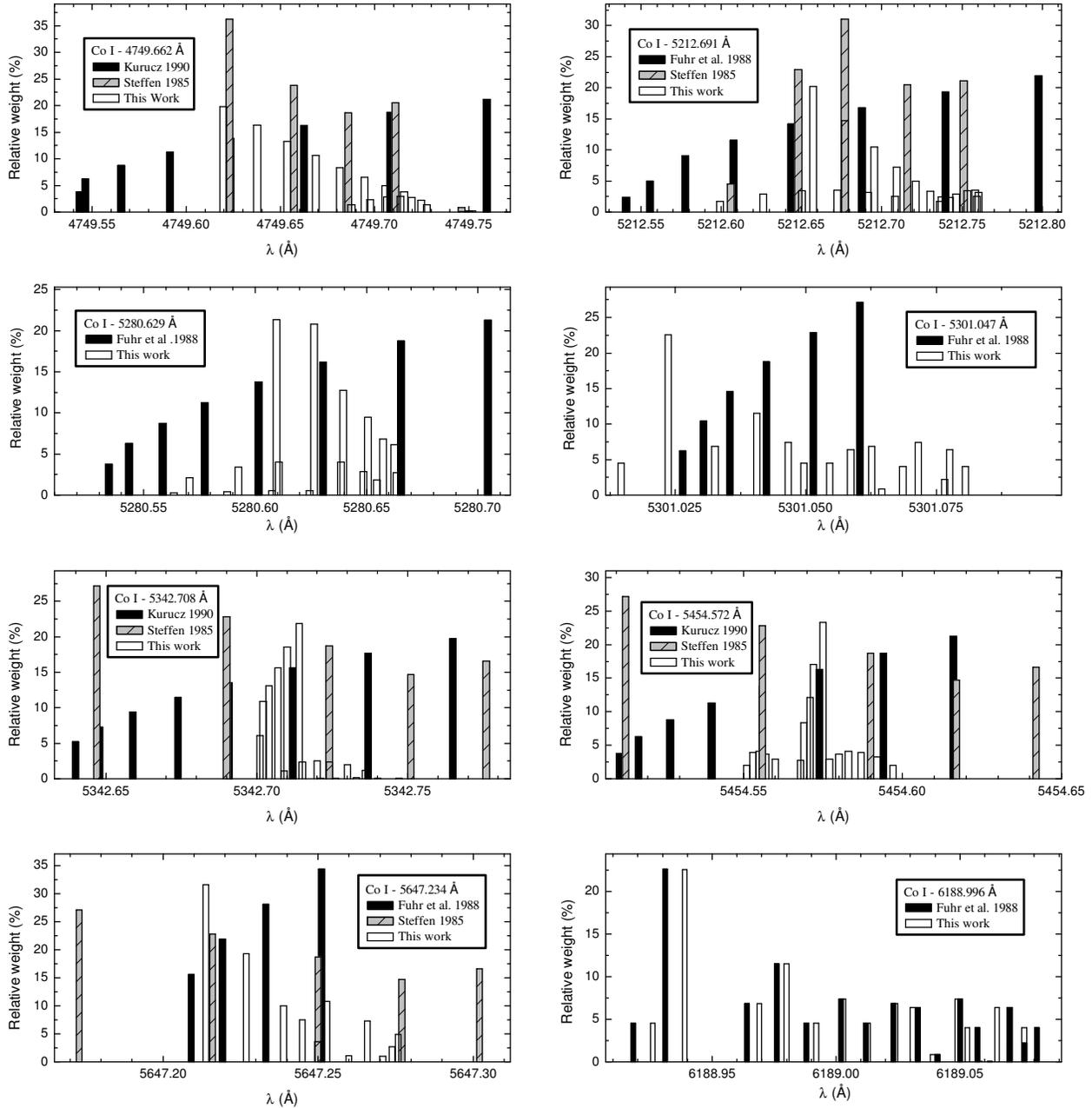}
\caption{HFSs of \ion{Co}{i} lines from KLL, S85, and our own
calculations. Note the dissimilarity of the structures (with the
exception of the line at 6188.996~\AA).}
\label{fig:hfs_co_sem_abund}
\end{figure*}

Of the 8 \ion{Co}{i} lines studied here, 5 have HFS data available
from both S85 and KLL (4749.662~\AA, 5212.691~\AA, 5342.708~\AA,
5454.572~\AA, and 5647.234~\AA). In
Fig.~\ref{fig:hfs_co_sem_abund} we compare these HFSs with those
calculated in this study. Note that we have employed newer, more
accurate laboratory values for the $A$ and $B$ interaction
factors, and our calculations are thus expected to be more
reliable than the previous ones. It can be seen that there is not
good agreement between the three HFS sets. This constitutes
further evidence of the very heterogeneous character of the S85
and KLL databases: while the HFSs of some of the lines contained
in those works have been very well calculated (like the
\ion{Mn}{i} line near 5400~\AA -- see
Sect.~\ref{sec:mn_line_by_line}), some present very strong
inconsistencies. For the other 3 \ion{Co}{i} lines, two of them,
at 5280.629~\AA\ and 5301.047~\AA, have HFSs from KLL that
disagree strongly with our calculations. However, for the Co line
at 6188.996~\AA\, there is good agreement between our calculations
and those from KLL. This offers yet another indication of the
heterogeneity of KLL: even coming from one single source
\citep{fuhretal88}, there are data with different levels of
accuracy.

\begin{figure}
\resizebox{\hsize}{!}{\includegraphics*{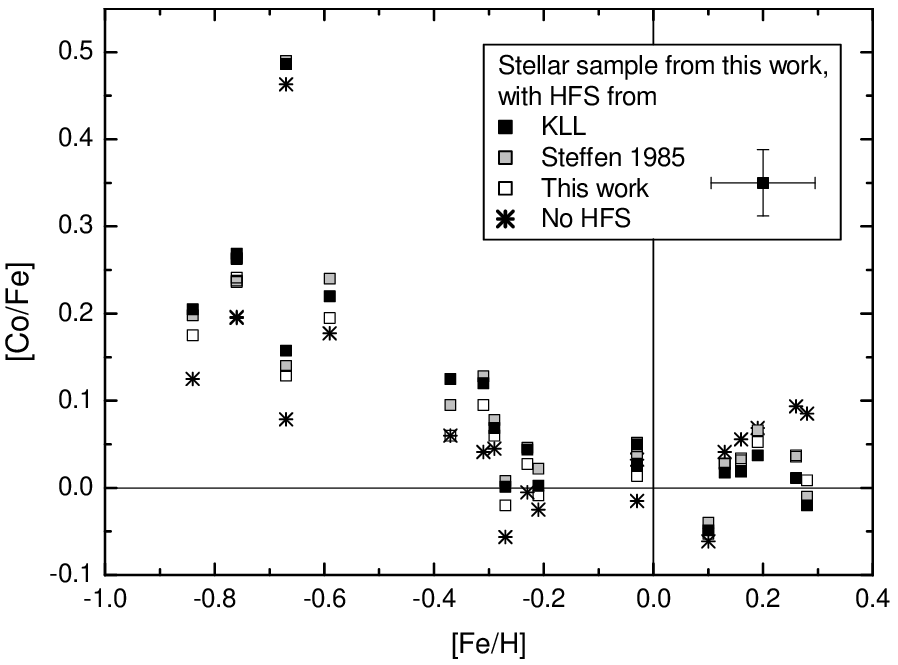}}
\caption{[Co/Fe] vs. [Fe/H] diagram displaying our sample star
results obtained without HFS and with three different sources of
HFS (our own calculations, S85 and the Kurucz's website, i.e.,
\citealt{kurucz90} and \citealt{fuhretal88}). There is good
agreement among all results, which indicates the small influence
of HFS on Co abundance determinations. An average error bar is
shown inside the legend box.} \label{fig:co_fe_fe_h_todos}
\end{figure}

A [Co/Fe] vs. [Fe/H] diagram containing our four sets of abundance
results, obtained without HFS and with the HFS data from S85, KLL,
and our own HFS calculations, is presented in
Fig.~\ref{fig:co_fe_fe_h_todos}. It is clear that the Co
abundances are not very sensitive to the inclusion of HFS: the
difference between determinations carried out with and without HFS
is at most 0.10~dex. This small influence of the HFS on the
derived abundances is a consequence of the small EWs of the
\ion{Co}{i} lines in our sample stars --
$\overline{\mbox{EW(\ion{Co}{i})}}=(23\pm15)~\mbox{m\AA}$, even if
we find quite pronounced differences in the HFS, as discussed
above. For cooler stars, with stronger \ion{Co}{i} lines, the
effect of HFS would be considerably more pronounced. The
differences in the average abundances obtained with HFS from
different sources are small, being at most 0.07~dex.

\section{Abundance trends and comparisons with results from the literature}

\subsection{Manganese}

\begin{figure}
\resizebox{\hsize}{!}{\includegraphics*{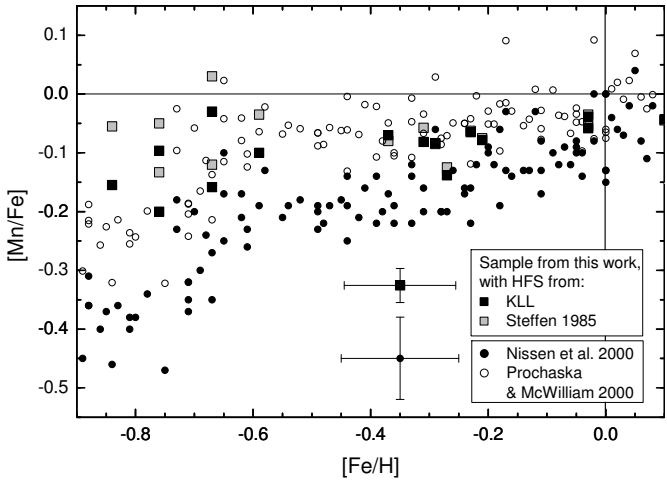}} \caption{
[Mn/Fe] vs. [Fe/H] diagram displaying our sample star results
obtained with two different sources of HFS -- S85 and KLL. Also
displayed are the objects from NCSZ00 and PM00. Both our sets of
results agree well with PM00, for $\mathrm{[Fe/H]}\gtrsim-0.5$.
For the more metal-poor objects, a good match is obtained between
our results using KLL HFS and PM00, whereas our results using S85
HFS tend to lie above PM00. Average error bars are displayed for
our data (large, full square) and for those of NCSZ00 and PM00
(small, full circle).} \label{fig:mn_fe_fe_h_todos}
\end{figure}

In Fig.~\ref{fig:mn_fe_fe_h_todos} we present a [Mn/Fe] vs. [Fe/H]
diagram which compares our derived manganese abundances (from the
HFS data in S85 and KLL) with the abundances obtained in the
studies of NCSZ00 and PM00, both of which include HFS in their
abundance calculations; NCSZ00 adopted HFS from S85, while PM00
used KLL. Inspection of this figure indicates the [Mn/Fe]
abundances derived in this study overlap well with the results
from PM00, but not with NCSZ00. NCSZ00 Mn abundances typically
fall below all other abundance results. This apparent
inconsistency can be explained by inhomogeneities in the database
of S85: the HFSs of the \ion{Mn}{i} lines used here (near
5400~\AA) seem to have been accurately calculated, agreeing well
with the HFSs from KLL, while those employed by NCSZ00 (near
6000~\AA) seem to present important discrepancies. As noted
before, HFSs for different lines, although from the same source,
can have quite different levels of reliability.

The origin of Mn has been associated with its production in SN~Ia
or SN~II, with the yields in SN~II being strongly metallicity
dependant. The Mn results from PM00, which were obtained from a
sample of 119 F and G~main-sequence stars from the thin disk,
thick disk, and the halo, indicated that SN~Ia's are the preferred
source for Mn mainly because the run of [Mn/Fe] versus [Fe/H]
showed a discontinuity at roughly $-0.7$ in [Fe/H]; this
metallicity representing the transition between the thin disk and
the thick dick/ halo. The overlap of our derived Mn abundances
with those from PM00 would support the idea that SN~Ia are
effective Mn producers, with no need to invoke production from
metallicity dependant yields in SN~II.

\subsection{Cobalt}

\begin{figure}
\resizebox{\hsize}{!}{\includegraphics*{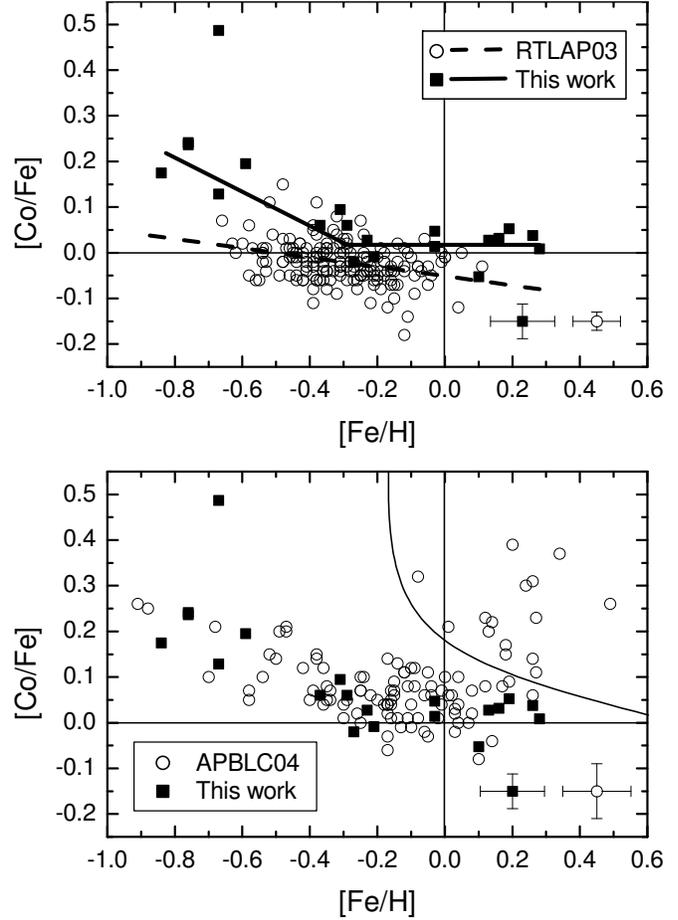}}
\caption{Comparison of our [Co/Fe] results with results from the
literature. Upper panel: RTLAP03. Average trends are shown as a
thick, solid line for our data and a thick, dashed line for
RTLAP03. Lower panel: APBLC03. A solid, curved line separates the
objects that seem to agree well with ours from those that do not.
In both panels, average error bars are shown in the lower right
corner.} \label{fig:co_literature}
\end{figure}

It is interesting to compare the cobalt abundance trends indicated by our
data with the results from other studies in the literature that
have analysed larger samples of stars. This is shown in
Fig.~\ref{fig:co_literature}, where we plot the [Co/Fe] abundances
from the study of \citealt{reddyetal03} (RTLAP03, top panel) and
\citealt{allendeprietoetal04} (APBLC04, bottom panel). The cobalt
results from this study in the figure are those which were obtained
with our HFS calculations.

Inspection of the top panel of Fig.~\ref{fig:co_literature}
indicates that the abundances obtained from our sample are roughly
coincident with the upper envelope of the RTLAP03 distribution
(for metallicities roughly between solar and $-0.4$). In fact, as
indicated by the lower metallicity stars in our sample and by the
thick line depicted in the figure, our Co abundances would seem to
exhibit a flat behaviour with nearly-solar values
($\overline{\mathrm{[Co/Fe]}}=+0.02\pm0.03$) for stars with
$\mathrm{[Fe/H]}\ge-0.30$, but increase linearly for the more
metal-poor objects, reaching $\mathrm{[Co/Fe]}=+0.22$ at
$\mathrm{[Fe/H]}=-0.80$. We note that RTLAP03 did not include HFS
in their Co abundance calculations and that they used EWs of three
lines, only one of which (at 5342~\AA) was used in this study.
Also, it seems that RTLAP03 tend to find [Co/Fe] lower than the
solar value for stars around solar metallicities.

APBLC04 determined their Co abundances from EWs of six lines,
three of which (at 5212~\AA, 5280~\AA, and 6188~\AA) were also
analyzed here. They did not employ HFSs. Their abundance pattern
is very similar to ours, seeming to exhibit the same flattening
for objects with $\mathrm{[Fe/H]}\ge-0.30$ and the same increase
for more metal-poor objects. One major difference, however, is
evident (although we have a much lower number of stars in our
sample): APBLC04 data exhibit a strong increase in the [Co/Fe]
abundance ratios for the objects with metallicities higher than
the Sun, with a large abundance dispersion. This behaviour has
also been reported by \citet{feltzing&gustafsson98} and
\citet{bodagheeetal03}. The discrepancy may be explained by the
lack of HFS in their analyses. Note that the large majority of the
stars with $\mathrm{[Fe/H]}>+0.1$ in APBLC04 have
$T_{\mathrm{eff}}\sim5000~\mathrm{K}$, resulting in stronger Co
lines. For this reason, the lack of HFS would lead to an incorrect
increase in the Co abundances of metal-rich objects. The authors
of stellar abundance analyses often take into account the HFS of
elements like Mn, Eu, and Ba, but usually neglect Co HFS, leading
to erroneous conclusions. We hope that, by virtue of the results
here presented, future studies will always include Co HFS in their
abundance determinations.

Comparing Fig.~\ref{fig:co_fe_fe_h_todos} to
Fig.~\ref{fig:mn_fe_fe_h_todos} we can see that the behavior of Co
with metallicity is clearly distinct from that of Mn. The origin
of cobalt, however, again involves production from both SN~II and
SN~Ia, with the relative contributions still uncertain. [Co/Fe]
rises from roughly solar metallicity to $\sim+0.2$, with a
behaviour that is reminiscent of an alpha-element.

\subsection{Europium}

\begin{figure}
\resizebox{\hsize}{!}{\includegraphics*{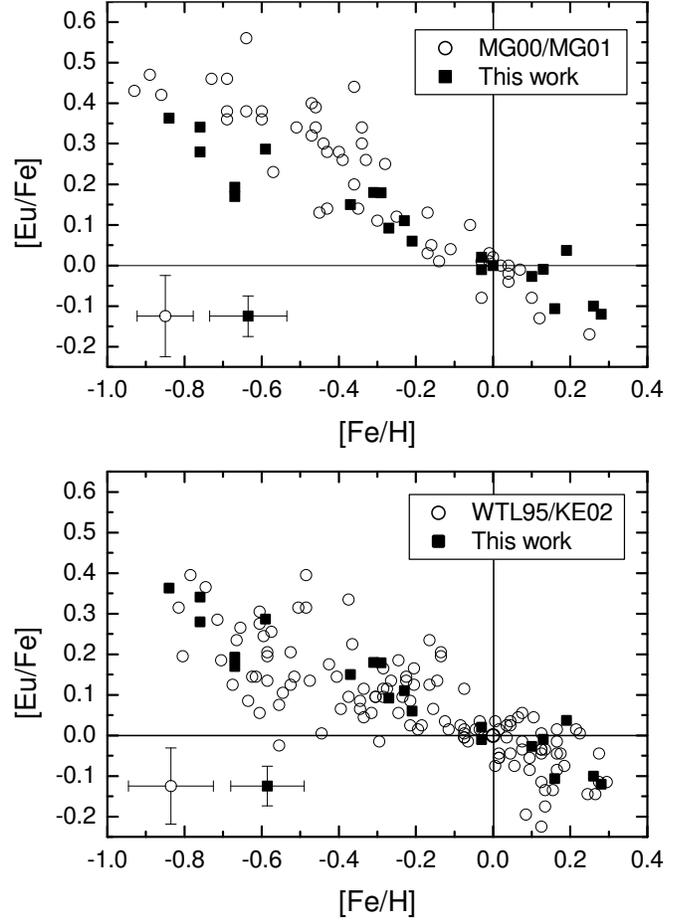}}
\caption{Comparison of our [Eu/Fe] results with the literature.
Upper panel: MG00/MG01. Lower panel: WTL95/KE02. The average error
bars from these works are displayed in the lower left corners of
both panels.} \label{fig:eu_literature}
\end{figure}

In Fig.~\ref{fig:eu_literature} we compare our [Eu/Fe] abundance
ratios to results from four other works: \citet[\space
MG00]{mashonkina&gehren00} and \citet[\space
MG01]{mashonkina&gehren01} -- upper panel; \citet[\space
WTL95]{woolfetal95} and \citet[\space KE02]{koch&edvardsson02} --
lower panel. Such a comparison is also of interest because it can
ultimately provide us with some additional check on our HFS
calculations.

MG00 and MG01 obtained Eu abundances for samples of halo and disk
stars, taking into account a non-local thermodynamical equilibrium
(NLTE) line formation. For comparison with our results, we have
retained only the disk stars with metallicities
$\mathrm{[Fe/H]}\ge-1.00$ and with accurate determinations of Eu
abundances (i.e., those not marked by ``:'' in their Tables). HFSs
were calculated by the authors using data from
\citet{beckeretal93} and \citet{brostrometal95}, as we did, but
they simplified the adopted structure by grouping close-by
components together, like S85. Spectral synthesis was employed to
analyse the same line we used. Our results agree well with theirs
for stars with $\mathrm{[Fe/H]}\ge-0.50$, but seem to be lower for
the more metal-poor objects.

The Eu abundances from WTL95 and KE02 were determined by spectral
synthesis using the same line we used, following a procedure
fundamentally identical to ours. The HFS used by WTL95 was taken
from \citet{krebs&winkler60}, who group close-by components
together, arriving at a total of 6~components per Eu~isotope. KE02
calculated their own HFS based on data taken from
\citet{brostrometal95}, also used here, retaining the complete,
detailed structure (16~components per isotope). NLTE effects are
minimal, because they are partially canceled out in the
differential analysis (as also happens in our work). KE02 merged
their database with that from WTL95 by means of a simple linear
conversion, obtained by intercomparison. Our abundances exhibit a
behaviour virtually identical to that of WLT95/KE02, but with
considerably lower scatter, as evident in the lower panel of
Fig.~\ref{fig:eu_literature}. Concerning the origin of Eu we refer
the reader to the thorough discussion of WTL95, where they
conclude that low-mass type~II supernovae are favoured as the main
\emph{r}-process site (97\% of all Eu is produced by the
\emph{r}-process, according to \citealt{burrisetal00}).

\section{Conclusions}

We present Mn, Co, and Eu abundances for a sample of 20 disk
dwarfs and subgiants of F5 to G8 MK spectral types with
$-0.8\le\mathrm{[Fe/H]}\le+0.3$. Our abundance trends for Mn with
metallicity are found to confirm the abundance results from
\citet{prochaska&mcwilliam00}, although both studies used
different sets of Mn I lines in the analyses, so this represents
an independent confirmation of the trend obtained in their study,
which favours type~Ia supernovae as the main astrophysical site of
Mn nucleosynthesis. In particular, our Mn results are in
disagreement with the trends previously found by
\citet{nissenetal00}, due to uncertainties in the HFS adopted in
their study. For Co, our results find a good agreement with the
trends with metallicity delineated by \citet{allendeprietoetal04}
for objects with $\mathrm{[Fe/H]}<0.0$, but significant
discrepancy is found for those with higher-than-solar metallicity.
The increase in Co abundances and high dispersion found by APBLC04
for the latter objects has been previously reported by
\citet{feltzing&gustafsson98} and \citet{bodagheeetal03}. We
believe this behaviour may be attributed to the lack of HFS in
their analyses. A comparison of our Co results with those by
\citet{reddyetal03} indicates that our Co abundances fall mostly
in the upper envelope of their distribution, for metallicities
lower than solar. The underabundance of their results may also be
connected to the lack of HFS in their analysis. Our Eu trend with
[Fe/H] was found to be in excellent agreement with other studies
in the literature (particularly with \citealt{woolfetal95} and
\citealt{koch&edvardsson02}).

In order to investigate the influence of HFS on the Mn and Co
abundances derived from our sample lines, we conducted
calculations with different HFS's from the literature, as well as
with HFSs calculated by us. For Mn, we find that for the four
\ion{Mn}{i} lines around 5400~\AA, the approximative HFS
calculations of S85 lead to nearly the same Mn abundances as
obtained with HFS from KLL. There are, however, large differences
in the Mn abundances calculated from the Mn I lines around
6000~\AA, as pointed out by \citet{prochaska&mcwilliam00}. The Co
abundances in this study (which were obtained from weak lines) are
weakly sensitive to HFS, presenting a 0.10~dex maximum difference
between determinations with and without HFS; they also are weakly
dependent on some details of the HFS calculations, such as small
variations between the selected $A$ and $B$ interaction factors
and grouping of close-by components. However, it is important to
note that the HFS's from different sources differ significantly
and the differences vary in magnitude for different \ion{Co}{i}
lines. These inconsistencies in the HFS data for different lines
reported here, would suggest that great care has to be taken when
considering the abundance of certain elements that require HFS
calculations.

\begin{acknowledgements}
We thank the referee for constructive criticism and comments that
led to a better paper. The authors wish to thank the staff of the
European Southern Observatory, La Silla, Chile. EFP acknowledges
financial support from CAPES/PROAP, FAPERJ/FP (grant
E-26/150.567/2003), and CNPq/DTI (grant 382814/2004-5). KC thanks
Andy McWilliam for helpful discussions. LS thanks the CNPq,
Brazilian Agency, for the financial support 453529.0.1 and for the
grant 301376/86-7. GFPM acknowledges financial support from
CNPq/Conte\'udos Digitais, CNPq/Institutos do Mil\^enio/MEGALIT,
FINEP/PRONEX (grant 41.96.0908.00), FAPESP/Tem\'aticos (grant
00/06769-4), and FAPERJ/APQ1 (grant E-26/170.687/2004).
\end{acknowledgements}

\bibliographystyle{aa}
\bibliography{referencias}

\end{document}